\documentclass{desyproc}

\usepackage{textcomp}
\usepackage[dvips]{graphicx}
\begin{document}
\title{Signatures for Solar Axions/WISPs}

\author{{\slshape Konstantin Zioutas$^{1,2}$, Theodoros Vafeiadis$^{1,2,3}$, Mary Tsagri$^1$\thanks{Present address: European Organization for Nuclear Research (CERN), CH-1211 Gen\`eve 23, Switzerland}, Yannis K Semertzidis$^4$, Thomas Papaevangelou$^5$, Theopisti Dafni$^6$\thanks{E-mail: tdafni@unizar.es} and Vassilis Anastassopoulos$^1$}\\[1ex]
$^1$University of Patras, Patras, Greece\\
$^2$European Organization for Nuclear Research (CERN), CH-1211 Gen\`eve 23, Switzerland\\
$^3$University of Thessaloniki, Thessaloniki, Greece\\
$^4$Brookhaven National Laboratory, NY-USA\\
$^5$IRFU, Centre d' \'{E}tudes Nucl\'eaires de Saclay, Gif-sur-Yvette, France\\
$^6$Laboratorio de F\'isica Nuclear y Astropart\'iculas, Universidad de Zaragoza, Zaragoza, Spain}



\maketitle

\begin{abstract}
Standard solar physics cannot account for the X-ray emission and other puzzles, the most striking example being the solar corona mystery. The corona temperature rise above the non-flaring magnetized sunspots, while the photosphere just underneath becomes cooler, makes this mystery more intriguing. The paradoxical Sun is suggestive of some sort of exotic solution, axions being the (only?) choice for the missing ingredient. We present atypical axion signatures, which depict solar axions with a rest mass m$_{ax}\simeq$17\,meV/c$^2$. Then, the Sun has been for decades the overlooked harbinger of new particle physics.

\end{abstract}

\section{Introduction}

The $\sim$5800\,K temperature of the solar photosphere naturally decreases outwards. Beyond a height of $\sim$500\,km, i.e., in the chromosphere, the beginning of a mysterious behaviour appears. The atmospheric temperature, instead of decreasing, starts rising again, up to 1-2\,MK, and this all of a sudden within only $\sim$100\,km. How this temperature behaviour can happen steadily, and all over the entire Sun, is dubbed the solar corona heating problem, {\it one of the most perplexing and unresolved problems in astrophysics to date} \cite{Ant10}. The phenomenology of the solar atmosphere makes this mystery even more enigmatic. For example, the unnaturally high-temperature upper solar atmosphere becomes even hotter above non-flaring magnetized locations, like the puzzling dark sunspots, reaching about 5-10\,MK \cite{Ziou09}, while the underlying surface of the sunspots gets relatively cooler, reaching occasionally $\sim$3000\,K, instead of the ambient $\sim$5800\,K. The temperature difference between two neighbouring solar layers, i.e., that of the photosphere and that of the corona, with the chromosphere sandwiched in between, widens. How can this additional and intriguing behaviour of the magnetized Sun fit conventional thinking? Obviously, the solar magnetic field, is the ingredient adding \textit{somehow} to the solar corona mystery. This finding is a second fingerprint of the corona's mystery, with the first being the formation of the surprisingly strong temperature inversion across the so-called transition region. Furthermore, following conventional reasoning, we still do not know how magnetic energy is converted into thermal energy of the corona \cite{War10}.

The solar corona mystery is not an isolated one, but rather ubiquitous throughout the solar-type stars in the Universe. Astonishingly, the Sun's radiation spectrum deviates strongly from that of a black body, and this reflects the whole mystery. For comparison, an almost perfect black body spectrum is exhibited by the CMB radiation of the infant Universe (3000\,K). Therefore, the question arises as to why the Sun behaves only partly as a perfect black body and how it manages to keep its tiny outer atmosphere, packed so close to its surface, at such an unnaturally high temperature. Note that the Sun is permeated spatiotemporally with unpredictably varying magnetic fields, which is the cause of many puzzling solar phenomena, while the early Universe had diminishingly small magnetic fields. This difference is essential from the axion point of view, since the axion-photon oscillation probability, as most solar phenomena, shows a striking B$^2$-dependence. One should bear in mind that no stellar theory expects a Sun-like star to emit any measurable quantity of X-rays, as we witness since decades with the Sun.

To the best of our knowledge, in recent times, no other solar problem has defied explanation for so long, e.g., take the solution of the solar neutrino deficit problem. It is logical to conclude that the mysterious coronal behaviour must be the manifestation of hidden new (solar) physics. Other solar phenomena associated with the mysterious 11-year clock, like flares, coronal mass ejections, sunspots, spicules, etc. raise also serious questions about their (not much less) mysterious origin, thus further suggesting a (common?) exotic solution.

\section{Solar/Stellar axion manifestation}
How can the solar behaviour be related to exotica like axions? The production of axions inside the Sun's core was widely accepted soon after their theoretical invention, constraining also their coupling strength to matter following the non-observation of additional star ageing effects \cite{Ziou09}. An extra energy escape from their hot core into space should have made them appear older than they actually are. In fact, stellar evolution arguments constrain the allowed escaping solar energy into axions to the \textperthousand-level. Nevertheless, this is still a quite large percentage compared to the solar observations under consideration as being due to or triggered by new exotica. For example, the unexpected quiet Sun X-ray emission makes only $\sim\!10^{-7}$  of its total energy output, while present X-ray missions detect solar fluxes at the level of $\sim\!10^{-14}$. This demonstrates the enormous potential solar observations have to unravel new physics, with the axion scenario inspiring the most, since (most) puzzling solar phenomena correlate with the magnetic field. Therefore, axion involvement in stars can have far reaching consequences, even if it causes only faint emission of radiation, since this leaves no signs of premature ageing.

Encouraged by the Sun's groundbreaking impact in the past in nuclear and astroparticle physics, it was natural to be attracted by the Sun's puzzling and inspiring behaviour, which depicts axion involvement. While we refer only to solar axions, the cause of the multifaceted and unpredictable Sun is not necessarily due to one single process by one single particle's involvement, though in certain cases exotic scenarios remain the only choice. Observations at extreme conditions, like the non-flaring quiet Sun during solar minimum or an isolated active/flaring solar region, could favour the showing-up of one exotic component against other(s), if any.

We refer throughout this work to axions, but we consider them as being representative of any other exotica dubbed WISPs (Weakly Interacting Slim Particles), which can couple similarly to ordinary matter, e.g., intriguing scalar particles like the chameleons, which are potential candidates for the cosmic dark energy. While the QCD-inspired axion implies a particle with one rest mass and one coupling constant, WISPs do not have to follow this constraint, e.g. massive solar axions of the Kaluza--Klein type \cite{DiL00}.

\section{Solar axion signatures}

For the last 15 years, the search for axions in solar X-rays has mostly been oriented towards a very light axion (rest mass$\,\ll\!10^{-4}$\,eV/c$^2$) \cite{Ziou09}. But, if its mass is (far) above this range, any search should fail, and this is the case so far. Therefore, this work addresses a much higher axion rest mass range ($\sim$20\,meV/c$^2$), albeit not intentionally but being observationally driven: if axions play a certain role in the Sun's workings, then some strange phenomena should show up, at least occasionally, i.e., with known physics being unable to provide an explanation. In fact, this is what happens so strikingly, since the Sun is full of mysteries.

We focus here on derived atypical solar axion signatures related to magnetic fields. For example, for the solar corona mystery, massive solar axions of the Kaluza--Klein type have been suggested as the potential source of the steady solar X-ray emission component \cite{DiL00}. The very thin solar corona is rather similarly hot ($\sim$1-10\,MK) as the hot solar core ($\sim$16\,MK), and therefore it requires an energy input, which has been elusive and has kept the corona mystery alive for several decades, even though there is no luck of proposed models. Note, the corona density changes dynamically, e.g., by factor up to $\sim$10-100 \cite{Asc04}. Then, within the axion scenario, the observed solar corona reflects the balance between inwards directed radiation pressure from the spontaneous decay of massive axions of the Kaluza--Klein type or any other massive, radiatively-decaying WISPs, while out-streaming axions being magnetically converted to X-rays exert an additional but outwards directed radiation pressure. To show the dynamical character of the Sun, it is worth mentioning, for example, the mysterious solar spicules, which cover about 1\% of the Sun's surface; their plasma density reaches values of $\sim\! 10^{11}$/cm$^3$, which are also of potential interest for axion-photon oscillations.

\begin{figure}[b]
 \centerline{\includegraphics[width=0.85\textwidth]{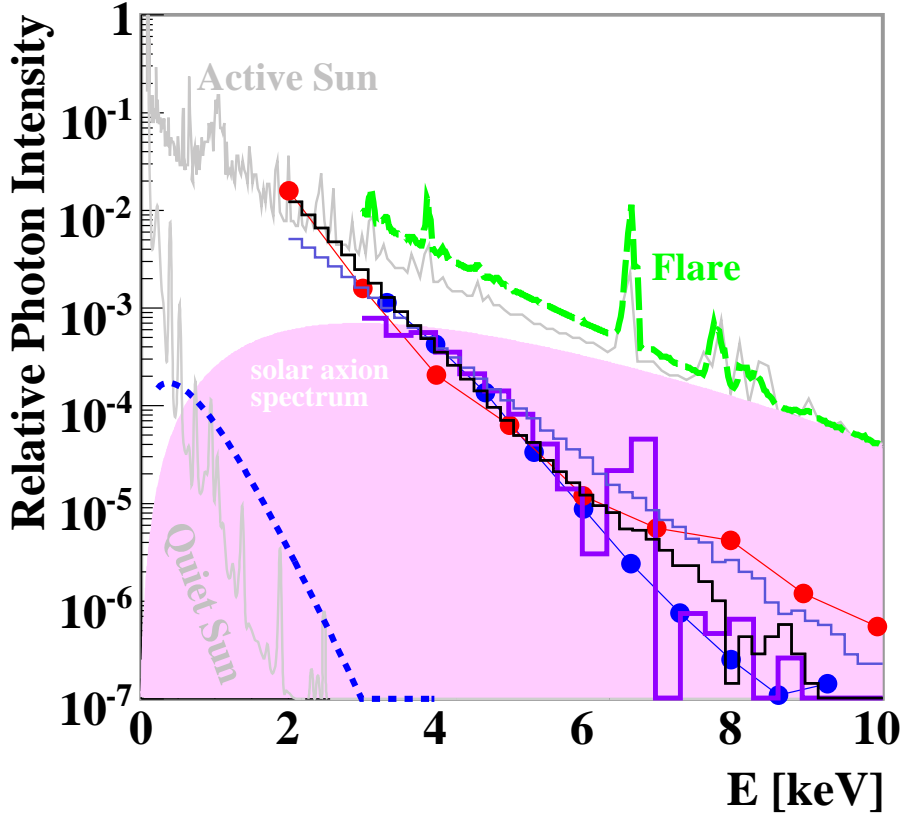}}
   \caption{Reproduced spectra from directly measured solar X-rays from: a) the main phase of a large flare with T$\sim$20\,MK \cite{Bat09} (green dashed), b) a flare with RESIK and RHESSI (red dots), c) preflaring periods after having subtracted the main X-ray flare component from the original spectra \cite{Bat09} (purple histogram), d) non-flaring active regions with T$\approx$6\,MK, i.e. sunspots (blue dots), e) non-flaring quiet Sun, with T$\approx2.7\,$MK, at solar minimum with SPhinX (blue dashed line) \cite{Syl10}; this is also supported by the recent findings that in the Quiet Sun regions stronger magnetic fields occur in deeper layers than in the ARs \cite{Ziou09}, implying more down-comptonization and giving rise to a larger slope. The initial broad solar axion spectrum is also shown shadowed (pink dashed line). Two GEANT4 simulated spectra following multiple Compton scattering from a depth of $\sim$350\,km and $\sim$400\,km are also shown for comparison (thin histograms), where the estimated plasma frequency, i.e., also the axion rest mass, is $\sim$17\,meV/c$^2$. The uncertainty is a factor of $\sim$2, since the density changes by factor of $\sim$4 between the $\sim$300\,km and $\sim$1000\,km depth. Note the strong deviations of the indirectly derived spectra (grey lines) in the past \cite{per00}  of the non-flaring quiet Sun at solar minimum and that of the active Sun at solar maximum below $\sim1\,$keV (SPhinX measurements). The spectra are not to scale.}
  \label{figure1}
\end{figure}

Figure \ref{figure1} shows the directly measured ``excess'' X-rays from the quietest to the flaring Sun (see Figure 10 in \cite{Ziou09}). Here we update the first intensity calculations presented in \cite{Ziou09}. Thus, a magnetic-field-related X-ray emission, be it transient be it steady, can be, in principle, axion in origin. The maximum conversion rate of out-streaming axions near the magnetized solar surface was estimated to be $P_{{\rm a}\to\gamma}\!\approx\!10^{-12}$ (see section 3 in \cite{Ziou09}). For comparison, assuming even that the entire quiet Sun soft X-ray luminosity measured recently by the SPhinX detector (L$_x\!\approx\!10^{21}$\,erg/s) is due to converted axions, this requires an even smaller conversion rate ($10^{-13}$). Furthermore, since only $\sim$1\% of the complex magnetism of the quiet Sun is seen \cite{Bue04}, this leaves room for much larger conversion efficiencies ($\sim10^{-8}$), i.e. an X-ray brightness of 10$^3$\,erg/cm$^2$/sec can still be axion related, which is not small. In addition, the fading solar magnetic field  during the 2009 solar minimum was correlated with a 100 times weaker soft X-ray emission than during the previous solar minimum measured by the SPhinX mission \cite{Syl10}. But, the only $\sim$25\% decrease of the solar magnetic field cannot justify a 100-fold X-ray luminosity decrease, following a B$^2$-dependence. But if, for example, the conversion occurs deeper inside the photosphere, some X-rays are absorbed, or, in any case they become more red-shifted and might evade observation. These measurements show that the calculated maximum axion conversion in \cite{Ziou09} was (very) conservative. Moreover, there is room for still larger conversion, which could account also for larger X-ray surface brightness from flares: the rarity of such events may eventually reflect the not so easily achievable `fine tuning' of magnetic field and plasma density. While the aim of this work is not to explain all solar X-ray phenomena exclusively by axions, this might be the case to a larger extent than anticipated so far (given the mentioned uncertainties).

\begin{figure}[hbt*]
  \centerline{\includegraphics[width=0.5\textwidth]{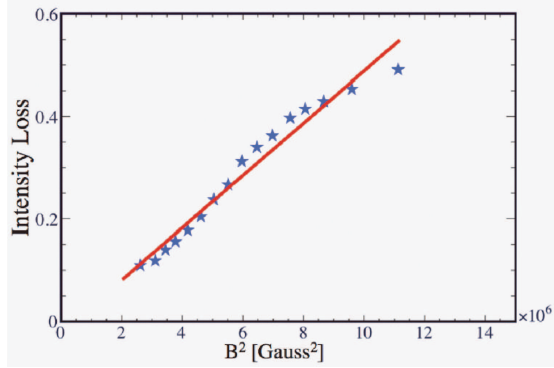}}
   \caption{The observed infrared (IR) intensity in the darkest position of sunspot cores (i.e., umbrae) is plotted vs. \textbf{B}$^2$, as  derived from a total number of 1392 sunspots \cite{Liv09,LivPP}. The measurements were performed from 1992 to 2009. The red line shows the \textbf{B}$^2$-dependence as a guide. It is not a fit to the IR intensity loss data. An example: intensity loss of 0.4 means that the number of IR photons is reduced by 40\% (with zero being the reference quiet Sun value) \cite{LivPP}. (Courtesy W. Livingston, NOAO/NSO, Tucson, Arizona.)}
  \label{figure2}
\end{figure}

\begin{figure}[hbt*]
  \begin{minipage}{0.49\textwidth}
      \centerline{\includegraphics[width=0.99\textwidth]
                                  {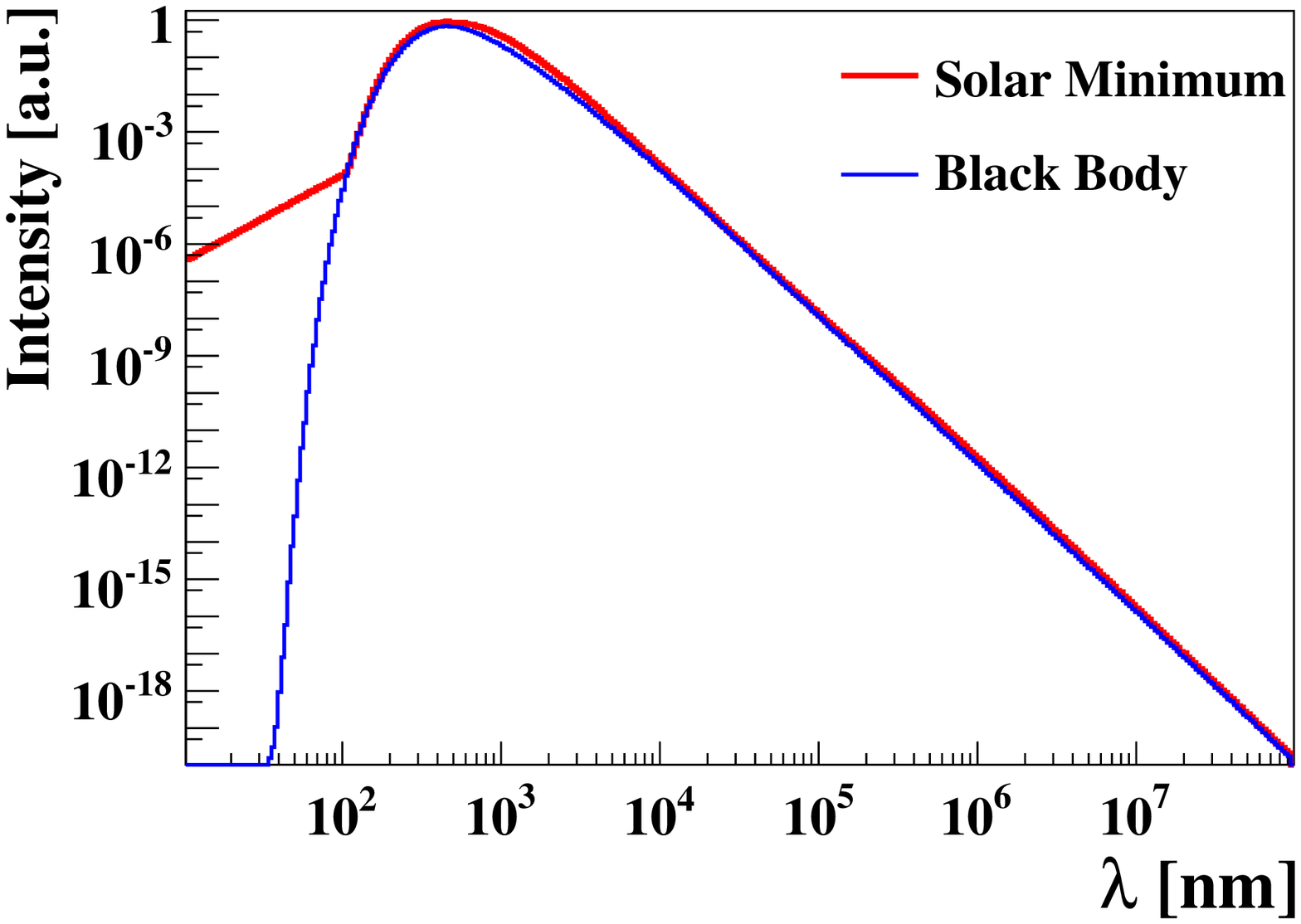}}
  \end{minipage}
  \begin{minipage}{0.49\textwidth}
      \centerline{\includegraphics[width=0.99\textwidth]
                                  {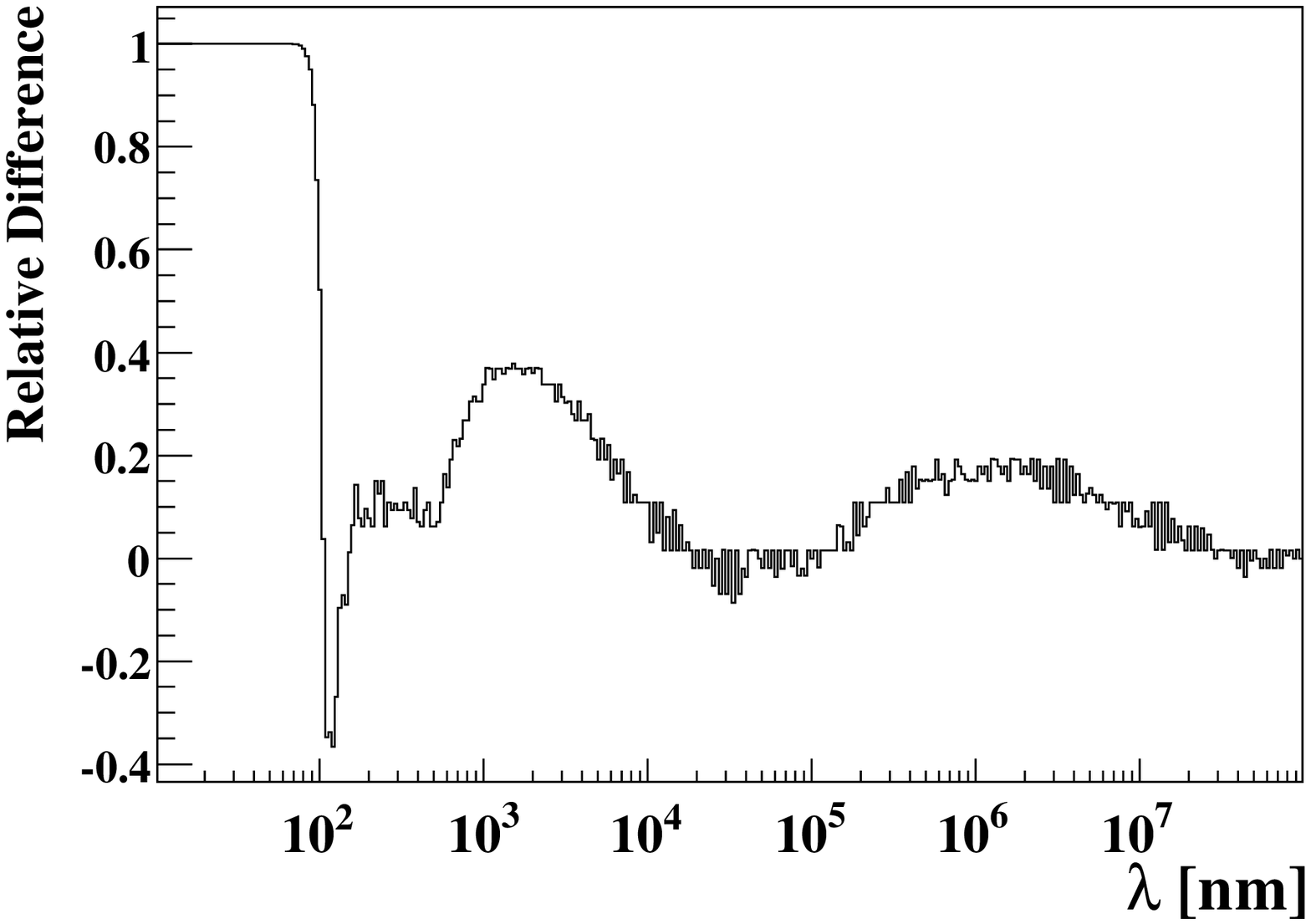}}
  \end{minipage}
   \caption{((\textbf{Left}) The approximated solar radiation spectrum during solar minimum (red line) and the black body spectrum for T=5800\,K (blue line). (\textbf{Right}) The normalized relative difference between the two spectra can be used to unravel non-thermal contributions, whatever their origin. The residuals below $\sim$100\,nm correspond to the hot corona excess. Residuals at $\sim$10$^6$\,nm might be a contamination of the CMB radiation, though the peak appears too broad towards shorter wavelengths to be eventually explained exclusively by CMB (either directly or reflected from the Sun). The origin of the excess around $\sim$2000--3000\,nm is real, but it is not yet identified. (Courtesy Marlene DiMarco/ UCAR Office of Education and Outreach/2009.)}
  \label{figure3}
\end{figure}

In addition, Figure \ref{figure2} shows the B$^2$ dependence of the  deficit IR emission above the magnetized  sunspots. If this is due to the disappearance  of  photons into axions, it implies also an overlooked strong solar axion source at low energies with far reaching implications in solar axion research. Finally, Figure \ref{figure3} explains how one may make visible new signatures, hidden in the solar irradiance spectrum, using the normalised residuals from a pure black body distribution.


\section{Conclusions}

We present observational evidence in favour of the solar axion scenario. Both massive and light axions are required, in order to explain the celebrated solar coronal heating mystery and unexpected (transient)  X-ray activity. The suggested axion scenario does not exclude the involvement of other WISPs (or the synergy with conventional phenomena). For example, the solar chameleon might be a potential candidate. This work is observationally driven. The accumulating axion signatures, when considered coherently all together, increase their combined significance in favour of solar axions as being at the origin of often miraculous solar behaviour. Nevertheless, each finding reflects an axion signature in its own right. On top of every other argument, we keep in mind that such a large amount of X-rays is anyhow not expected to be emitted by a cool star like the Sun, and this is what triggered this work.



\begin{footnotesize}

\end{footnotesize}



\begin{thebibliography}{99}
\bibitem{Ant10}
P.~Antolin, K.~Shibata, T.~Kudoh, D.~Shiota and D.~Brooks, ``Magnetic Coupling between the Interior and Atmosphere of the Sun,'' Astroph.\ Space Sc.\ Proc.\ Part 2  pp.277-280 (2010) [eprint astro-ph/0903.1766].

\bibitem{Ziou09}
K.~Zioutas, M.~ Tsagri, Y.~K.~Semertzidis, T.~ Papaevangelou, T.~ Dafni and V.~ Anastassopoulos, ``Axion Searches with Helioscopes and astrophysical
signatures for axion(-like) particles,'' New J.\ Phys.\ {\bf 11} 105020 (2009) and references therein.

\bibitem{War10}
H.~P.~Warren et al., ``Evidence for Steady Heating: Observations of an Active Region Core with Hinode and TRACE,'' ApJ {\bf 711} p.228 (2010).

\bibitem{DiL00}
L.~DiLella, P.~Pilaftsis, G.~Raffelt and K.~Zioutas, ``Search for solar Kaluza-Klein axions in theories of low-scale quantum gravity,'' Phys.\
Rev.\ D {\bf 62} 125011 (2000);
 L.~Di~Lella and K.~Zioutas ``Observational evidence for gravitationally trapped massive axion (-like) particles,''
 Astropart.\ Phys.\ {\bf 13} p.145 (2003).

\bibitem{Asc04}
M.~Aschwanden, ``Physics of the Solar Corona,'' Ed. Springer pp.24-26 (2004).

\bibitem{Bue04}
J.~T.~Bueno, N.~Shchukina and A.~A.~Ramos, ``A substantial amount of hidden magnetic energy in the quiet Sun ,'' Nature {\bf 430} p.326 (2004).

\bibitem{Bat09}
M.~Battaglia, L.~Fletcher, A.~O.~Benz, ``Observations of conduction driven evaporation in the early rise phase of solar flares,'' A.\& A. {\bf 498} 891 (2009).

\bibitem{Syl10}
J.~Sylwester, M.~Kowalinski, S.~Gburek, M.~Siarkowski, S.~Kuzin, F.~Farnik, F.~Reale, K.~J.~H.~Phillips, ``The Sun's X-ray Emission During the Recent Solar Minimum,'' Eos Trans.\ AGU, {\bf 91 }(\#8) p.73 (2010).

\bibitem{per00}
G.~Peres, S.~Orlando, F.~Reale, R.~Rosner and H.~Hudson ``The Sun as an X-Ray Star. II. Using the Yohkoh/Soft X-Ray Telescope-derived Solar Emission Measure versus Temperature to Interpret Stellar X-Ray Observations,'' ApJ. {\bf 528} 537 (2000).

\bibitem{Liv09}
 W.~Livingston, M.~Penn, ``Are Sunspots Different During This Solar Minimum?,'' Eos Trans. AGU {\bf 90} (\#30) p.257 (2009)
                       http://www.leif.org/EOS/2009EO300001.pdf.

\bibitem{LivPP} W.~Livingston, private communication , 2010.

\end{thebibliography}
\end{document}